\documentclass[twocolumn,showpacs,preprintnumbers,amsmath,amssymb,prl,superscriptaddress]{revtex4-1}
\usepackage{graphicx}
\usepackage{dcolumn}
\usepackage{bm}
\usepackage{color}
\usepackage[colorlinks=true, citecolor=blue, urlcolor=blue]{hyperref}

\newcommand{\Nc}{N_{\text{c}}}
\newcommand{\LQCD}{\Lambda_{\text{QCD}}}
\newcommand{\muB}{\mu_{\text{B}}}
\newcommand{\muq}{\mu_{\text{q}}}
\newcommand{\rhoB}{\rho_{\text{B}}}
\newcommand{\Tpc}{T_{\text{pc}}}
\newcommand{\Tch}{T_{\text{ch}}}
\newcommand{\MeV}{\text{\,MeV}}
\newcommand{\GeV}{\text{\,GeV}}
\newcommand{\fm}{\text{\,fm}}
\newcommand{\bB}{\boldsymbol{B}}
\newcommand{\bomega}{\boldsymbol{\omega}}
\newcommand{\hypertriton}{{}^3_\Lambda{\rm H}}
\newcommand{\atmark}{\textcircled{\it a}}

\begin{document}
\title{Little-Bang and Femto-Nova in Nucleus-Nucleus Collisions} 
{
\author{Kenji Fukushima}
\affiliation{Department of Physics, The University of Tokyo,
  7-3-1\! Hongo,\! Bunkyo-ku, Tokyo\! 113-0033, Japan}
\author{Bedangadas Mohanty}
\affiliation{School of Physical Sciences, National Institute of
  Science Education and Research, HBNI, Jatni 752050, India}
\affiliation{Institute of Modern Physics, Chinese Academy of Sciences, 
  509 Nanchang Road, Lanzhou 730000, China}
\author{Nu Xu}
\affiliation{Institute of Modern Physics, Chinese Academy of Sciences, 
  509 Nanchang Road, Lanzhou 730000, China}
\affiliation{College of Physical Science and Technology, Central China
  Normal University, Wuhan 430079, China}
\affiliation{Nuclear Scince Division, Lawrence Berkeley National
  Laboratory, Berkeley, CA 94720, USA}

\date{\today}
\begin{abstract}
  We make a theoretical and experimental summary of the
  state-of-the-art status of hot and dense QCD matter studies on
  selected topics.  We review the Beam Energy Scan program for the QCD
  phase diagram and present the current status of search for QCD
  Critical Point, particle production in high baryon density region,
  hypernuclei production, and global polarization effects in
  nucleus-nucleus collisions.  The available experimental data in the
  strangeness sector suggests that a grand canonical approach in
  thermal model at high collision energy makes a transition to the
  canonical ensemble behavior at low energy.
  We further discuss future prospects of nuclear collisions to
  probe properties of baryon-rich matter.  Creation of a quark-gluon
  plasma at high temperature and low baryon density has been called
  the ``Little-Bang'' and, analogously, a femtometer-scale explosion of
  baryon-rich matter at lower collision energy could be called the
  ``Femto-Nova'', which may possibly sustain substantial vorticity and
  magnetic field for non-head-on collisions.
\end{abstract}
\maketitle

\section*{INTRODUCTION}

Nuclei are bound states of protons and neutrons via the Strong
Interaction and the fundamental theory of the Strong Interaction is
Quantum Chromodynamics (QCD).  In QCD gluons are
massless due to gauge symmetry and up ($u$) quarks are as light as
$m_u=3-4\MeV$ and down ($d$) quarks are heavier than $u$-quarks,
i.e., $m_u/m_d\sim 0.5$ (see Ref.~\cite{Sanfilippo:2015era} for a
recent review of quark masses from the lattice-QCD).  The strange
($s$) quark mass is comparable to the typical QCD energy scale; that
is, $m_s=80-90\MeV$ of the same order as $\LQCD=100-200\MeV$.  Since
charm ($c$) and bottom ($b$) quarks are much heavier than $\LQCD$,
they would make only small contributions to bulk thermodynamics and
they serve as external probes.  Here, we focus on two puzzling QCD
features for the nucleons which are composed of $\Nc$ valence quarks
(where $\Nc=3$ is the number of colors in QCD) and have a mass,
$m_N\simeq 940\MeV \sim \Nc\LQCD$.  The first question is; how can
almost massless particles form a bound state with a positive binding
energy?  The second question is; how can the nucleons become such
extremely massive out of almost massless particles?  The former question on
the existence of bound states is referred to as \textit{confinement}
and the latter on the origin of the mass is via
\textit{spontaneous chiral symmetry breaking}.

The key to resolve these puzzles lies in the QCD vacuum structure.
The vacuum in quantum field theory is not empty in general, but is
full of  quantum fluctuations dictated by fundamental interactions.
Thus, the QCD vacuum is regarded as a ``medium'' in analogy to
condensed matter physics.  Just like spin systems for example, the
QCD vacuum structure may be either an ordered/disordered state
according to external environments such as the temperature $T$, the
baryon density $\rhoB$ (or the baryon chemical potential $\muB$), the
magnetic field $\bB$, the vorticity $\bomega$, etc.  The idea of the
relativistic nucleus-nucleus collisions or heavy-ion collisions is to
shake the QCD vacuum with high energy density to observe new states of
matter out of quarks and gluons and to seek for traces of phase
transitions associated with confinement and/or chiral symmetry
breaking (see Ref.~\cite{Baym:2001in} for historical backgrounds).

The saturation density around the center of heavy nuclei is
$\rho_0\simeq 0.16\,\text{nucleons$/$fm$^3$}$ corresponding to
the energy density $\epsilon_0 \simeq 0.15\GeV/\text{fm}^3$.  At the
initial stage of the heavy-ion collisions the energy density can reach
hundreds times larger than the saturation density depending on the
collision energy $\sqrt{s_{NN}}$ per nucleon.  The Stefan-Boltzmann
law converts the energy density $\epsilon = 0.8-1.0\GeV/\text{fm}^3$ to
$T=150-160\MeV$ that is comparable to $LQCD$, above which a
quark-gluon plasma (QGP) is realized.  In the history of the Universe
such high-$T$ matter should have existed shortly (of the order of $\mu$s) after 
the Big Bang.  In this sense the QGP physics can be regarded as an
emulation of the Big Bang in the laboratory on the Earth,
which may well be called the \textit{Little Bang}.

\begin{figure}
  \includegraphics[width=0.9\columnwidth]{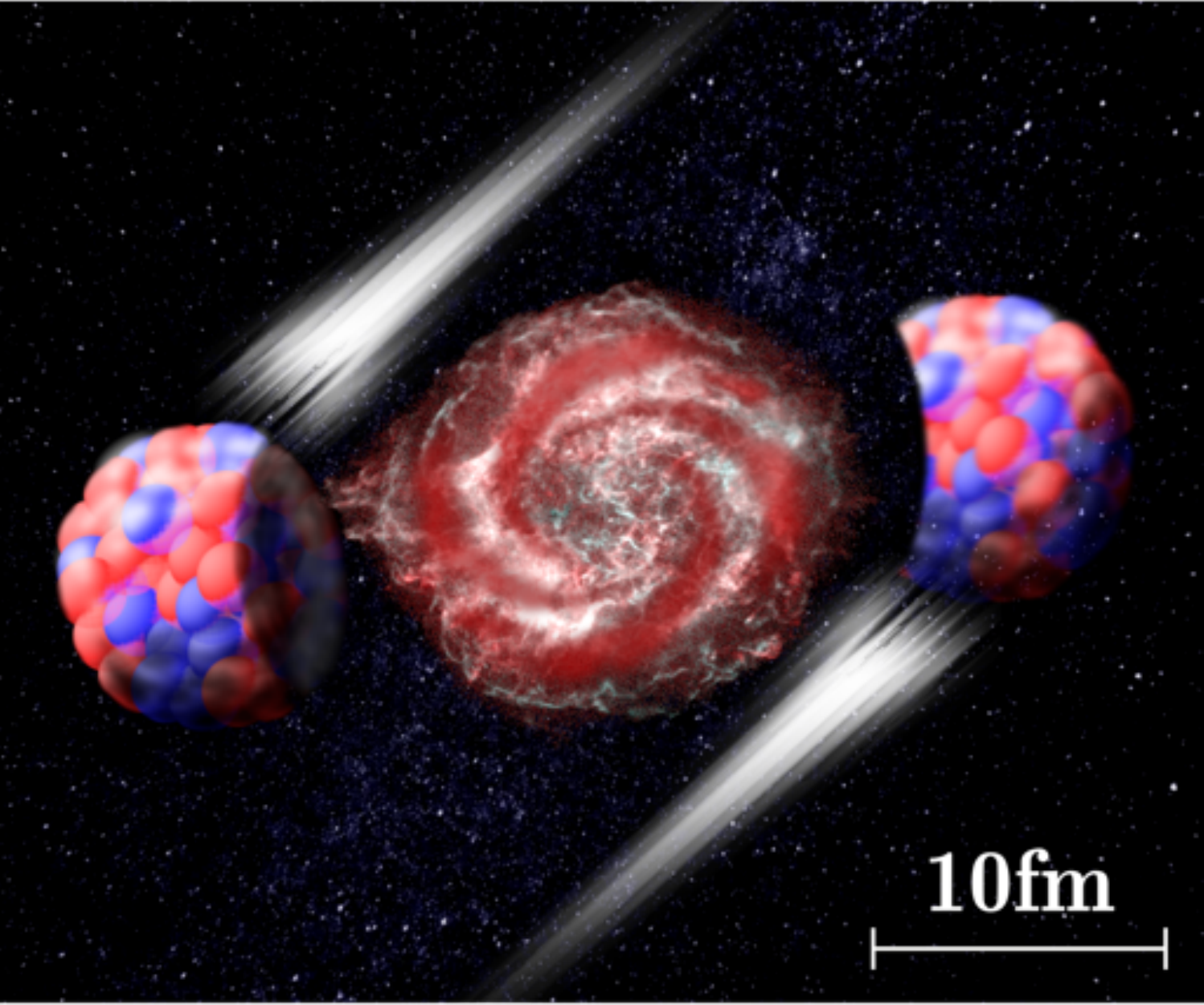}
  \caption{Schematic illustration of Femto-Nova as a femtometer-scale 
    explosion of baryon-rich matter in relativistic heavy-ion collisions.}
  \label{fig:FemtoNova}
\end{figure}

Now that we have learnt intriguing properties of the QGP in the regime
at high $T$ and low $\muB$, a next direction of the relativistic heavy-ion
collision physics is expanding toward the regime at high baryon
density.  The phase structures could have much richer contents in
these high-density regions, see
Refs.~\cite{Fukushima:2013rx,Buballa:2014tba,Fischer:2018sdj} for
reviews on the QCD phase diagrams.  We have already known that
symmetric nuclear matter has a first-order phase transition at
$\muB\approx 923\MeV$; that is, the nucleon mass minus the binding
energy $\sim 16\MeV$.  Then, a terminal point of the first-order phase
boundary, namely, a critical point of liquid-gas phase transition in
nuclear matter exists around
$T= (10-20)\MeV$ (see Ref.~\cite{Chomaz:2004nw} for a
comprehensive review including experimental signatures).  The
question is what new state of matter is anticipated for nuclear matter
at higher baryon density.  The central cores
of the neutron star would exhibit the most baryon-rich and equilibrated
state of matter in the Universe, where the density could be as large
as $\sim 5\rho_0$ or even higher.  The baryon density may become even
larger in transient states such as from neutron star mergers.
Analogously, by adjusting the collision energy in the heavy-ion
collisions, the baryon density could transiently increase up to
several times $\rho_0$ according to numerical
simulations~\cite{Arsene:2006vf}.  Such a femtometer-scale explosion
of baryon rich matter may be called the \textit{Femto-Nova}.
This concept is illustrated in Fig.~\ref{fig:FemtoNova}.

So far, most of the high-energy nuclear collisions have been designed
at high energy region where the temperature of the formed matter is
high in order to assure the formation of the QGP.
Unfortunately, only little is known from theory about the phase
structures in low-$T$ and high-$\muB$ (or high-$\rhoB$) regions.
There are some speculations on the ground state structures in such
regimes which will be partially reviewed in this article.  The most
important landmark is the \textit{QCD Critical Point}.  In the same
way as the critical point of the nuclear matter liquid-gas phase
transition, deconfined QCD matter may have
a first-order phase boundary and the QCD Critical Point appears at the
terminal point of the first-order line.  Its exact location is not
well constrained yet, and the experimental efforts to discover the QCD
Critical Point are still continuing.

\section*{CURRENT STATUS -- THEORY} 

\paragraph*{Theoretical Background for the QCD Phase Diagram:}
It has been established that hadronic matter continuously changes into
the QGP with increasing temperature $T$ as long as $\muB$ is sufficiently
smaller than $T$.  Although this continuous
change of matter takes place around
$T=(156.5\pm1.5)\MeV$ according to the lattice-QCD
simulation~\cite{Bazavov:2018mes}, there is no strict phase
transition.  In the QCD community this transitional but still
continuous change of matter is commonly referred to as a
{\it crossover}.

If masses of $u$ and $d$ quarks are zero and other quarks are massive,
such 2-flavor QCD matter would go through a second-order phase
transition for which various derivatives of thermodynamic quantities
diverge with critical exponents belonging to the O(4) universality
class (or a first-order transition if the axial anomaly is restored,
see Ref.~\cite{Pisarski:1983ms}).  Physical values of $m_u$ and $m_d$
are much smaller than
$\LQCD$, and it is conceivable to observe some remnants of the
second-order phase transition.  Indeed, in the lattice-QCD
simulation~\cite{Ejiri:2009ac}, it has been numerically confirmed that
the magnetic scaling follows consistently with the O(4) universality
class, from which the ``pseudo-critical'' temperature, $\Tpc$, can be
deduced.

The pseudo-critical temperature should be a function of the density.
Generally $\Tpc(\muB)$ is a decreasing function
with increasing $\muB$ due to the Pauli blocking of quarks in phase space.
Because of the notorious sign problem in the Monte-Carlo algorithm,
expectation values of observables cannot be computed and
the first-principles lattice-QCD simulation cannot access a region
with a substantial value of baryon density, and so there is no
reliable theoretical prediction for $\Tpc(\muB)$ at $\muB$ much larger
than $T$.

Instead, a phenomenologically determined boundary on the $\muB$-$T$
plane is known, called the line of the {\it Chemical Freeze-out},
which has been identified from a
hybrid approach of theory and experiment.  The Chemical Freeze-out
literally means that inelastic reactions stop and chemical compositions
are fixed there.  In QCD matter hadrons interact and particle
species can change like chemical reactions.  In the heavy-ion
collision the physical system is expanding and the temperature rapidly
drops down.  Therefore, the average inter-particle distance would
increase as the temperature gets lowered.  In particular, around
$\Tpc$, the entropy density significantly falls down, and
correspondingly the degrees of freedom in thermal excitations
decrease~\cite{BraunMunzinger:2003zz}.  Since the average inter-particle
distance becomes large, the interaction among hadrons is
considered to diminish.  In this way the observed particle yields keep
the footprint of hot and dense hadronic matter when the interaction
was turned off, that is presumably when the matter underwent a
crossover at $\Tpc$.  If this scenario is the case, the particle
yields should be sensitive to $\Tpc(\muB)$.  As observed experimentally,
relative abundances of hadrons obey the thermal distribution at common
$T$ and $\muB$,  so that the thermal fit can fix $T$ and
$\muB$, or a line of Chemical Freeze-out,
$T=\Tch(\muB)$~\cite{Andronic:2017pug}.  We note that
the charge chemical potential $\mu_Q$ is fixed from the proton/neutron
ratio and the strangeness chemical potential $\mu_S$ is fixed from the
strangeness free condition.  With various center-of-mass colliding
energies, we can change accessible $\muB$ to sample $\Tch(\muB)$ from
the thermal fit~\cite{Adamczyk:2017iwn, Gupta:2020pjd}, see
Fig.~\ref{fig:CF}.  Generally speaking, collisions at smaller
$\sqrt{s_{NN}}$ have larger baryon stopping, leading to larger values
of $\muB$ (and smaller values of $T$)~\cite{Cleymans:2005xv}.  Thus,
the line of Chemical Freeze-out on the $\muB$-$T$ plane should be
regarded as an experimentally determined QCD phase diagram which
appears to be consistent with the lattice-QCD
results at the vanishing $\muB$ region~\cite{Bazavov:2018mes}, as also
displayed in Fig.~\ref{fig:CF}.  This underlies the idea of the Beam
Energy Scan (BES) program at RHIC{}.  The thermal description is
applicable for not only particle abundances but also thermodynamic
quantities such as the pressure and the entropy density.
\vspace{1em}

\begin{figure}
  \includegraphics[width=0.95\columnwidth]{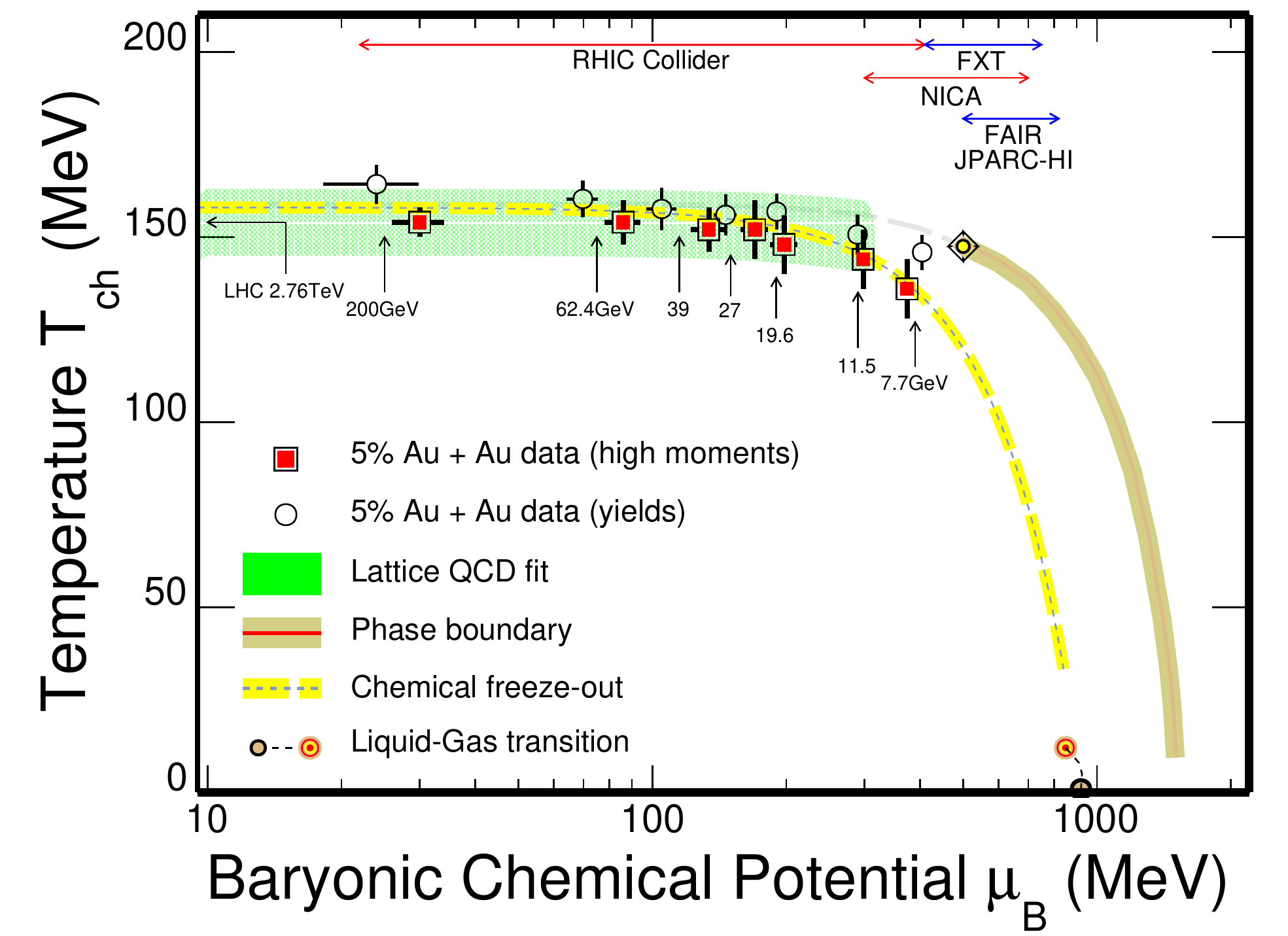}
  \caption 
  {Chemical Freeze-out temperature $\Tch(\muB)$ from the top 5\%
    central Au+Au collisions at RHIC{}.  Open-circles represent the 
    parameters extracted from hadron yields~\cite{Adamczyk:2017iwn}, 
    while the filled-squares are extracted from net-proton higher 
    moments (up to third order)~\cite{Gupta:2020pjd}.  Representing the 
    smooth-crossover region are the lattice-QCD results shown as 
    green-band.  The empirical thermal fit results to global hadron yield 
    data are shown as yellow-line~\cite{Cleymans:2005xv}.  The coverage 
    of the RHIC BES program, STAR fixed target program (FXT), and future 
    (FAIR, JPARC-HI, and NICA) experimental facilities are also 
    indicated at the top of the figure.  The liquid-gas transition 
    region that features a second order critical point is shown by the 
    red-circle, and a first-order transition line is shown by the 
    black dashed line, which connects the critical point to the ground 
    state of nuclear matter.}
  \label{fig:CF}
\end{figure}

\paragraph*{Observables for the QCD Critical Point:}
It is known from the theoretical analysis that the QCD crossover has a
general tendency to become closer to a first-order transition at
larger $\muB$ (see discussions
in a review~\cite{Fukushima:2010bq}).  It is thus a natural
anticipation that the QCD crossover may turn to a first order phase
transition in a high-density regime.  If this is the case, as
suggested by some effective model studies, there must be a
``critical'' value of $\muB$ above which a first-order phase
transition occurs and below which only a crossover is found.  This
separating point is the QCD Critical Point and critical
fluctuations associated with the second-order phase transition should
be expected at this point.  Its exact location is still under dispute,
and the lattice-QCD results~\cite{Bazavov:2017dus} disfavors the
existence of the QCD Critical Point for $\muB/T \lesssim 2$.

Interestingly, the QCD Critical Point emerges with nonzero physical
quark masses, so that it belongs to not the O(4) but the Z(2)
universality class.  Moreover, the dynamical universality class has
been also identified as the model H (dynamics of the liquid-gas
critical point of a fluid)~\cite{Son:2004iv} (see a
review~\cite{RevModPhys.49.435} for detailed classification).
The dynamical critical exponents are important inputs for simulations
including the critical slowing down effects~\cite{Berdnikov:1999ph}.

For experimental signatures, we can in principle seek for enhanced
fluctuations coupled to the critical modes.  Since the critical modes
appear in a mixed channel of scalar (i.e., chiral condensate) and
vector (i.e., baryon density) at the QCD Critical
Point~\cite{Fujii:2003bz}, the baryon number fluctuations are
sensitive to the criticality.  Let us denote the baryon number
fluctuation by $\delta N= N - \langle N\rangle$ where $N$ is the
number of net baryons at each collision event and
$\langle \cdots \rangle$ stands for the ensemble average taken over
collision events.  At the critical point, generally, the correlation
length $\xi$ diverges, and it was pointed out in
Ref.~\cite{Stephanov:2008qz} that the non-Gaussian fluctuations behave
as
\begin{equation}
  \langle (\delta N)^k \rangle_c \;\sim\;
  \xi^{k(5-\eta)/2 - 3} \,.
  \label{eq:nonGauss}
\end{equation}
Here, the subscript $c$ represents a part of the correlation function
corresponding to the connected diagrams (to extract non-Gaussian
fluctuations) and $\eta$ is the anomalous dimension (which is
usually $\eta\ll 1$).  Higher-order fluctuations are more
sensitive to the criticality, but they need more statistics especially
to construct connected contributions.  Now, the third order ($k=3$)
and the fourth order ($k=4$) in Eq.~\eqref{eq:nonGauss} are common
measures for the QCD Critical Point search,
normalized by the $k=2$ fluctuation (variance),
$\sigma^2=\langle (\delta N)^2\rangle$; namely,
\begin{equation}
  S = \frac{\langle (\delta N)^3\rangle}{\sigma^3}\,,\qquad
  \kappa = \frac{\langle (\delta N)^4 \rangle_c}{\sigma^4}
  = \frac{\langle (\delta N)^4\rangle}{\sigma^4} - 3\,.
  \label{eq:Nfluct}
\end{equation}
$S$ and $\kappa$ are called the skewness and the kurtosis and
characterize how skewed and how sharp the distribution of $\delta N$
appears, respectively.  It is noted that, in the QCD-physics context,
$\kappa$ (or the fourth-order cumulant) was first considered in the
lattice-QCD simulation to diagnose whether quarks are confined or
deconfined~\cite{Ejiri:2005wq}.

This idea can be easily generalized to other observables coupled to
the critical modes.  Because observables in the heavy-ion collisions
are integrated quantities over the whole dynamical evolutions, we
should look at fluctuations of conserved charges;  otherwise, critical
enhancement would be wiped off through the dynamical evolution.  There
are three representative candidates available in the heavy-ion
collisions, i.e., the baryon number ($B$), the electric charge ($Q$),
and the strangeness ($S$).  In thermodynamics those fluctuations are
defined by the derivatives of the pressure with respect to the
chemical potentials corresponding to conserved charges,
i.e.~\cite{Karsch:2010ck}
\begin{equation}
  \chi_q^{(n)} = \frac{\partial^n [ p(T,\muB,\mu_Q,\mu_S)/T^4]}
  {\partial(\mu_q / T)^n} \,,
\end{equation}
where $q=B,\, Q,\, S$.  In terms of these fluctuations the skewness
and the kurtosis are represented as
$S_q \sigma_q = \chi_q^{(3)}/\chi_q^{(2)}$ and
$\kappa_q \sigma_q^2 = \chi_q^{(4)}/\chi_q^{(2)}$, respectively.
Susceptibilities in mixed channels can also be defined in a similar
fashion.

The baseline to be compared for the critical enhancement is estimated
by an approximation of non-interacting and dilute hadronic gases
described by the Boltzmann distribution.  Then, in this Boltzmann gas
approximation, the chemical potential dependence is factored out,
yielding,
\begin{equation}
  S_q \sigma_q \simeq \tanh(\mu_q/T)\,,\qquad
  \kappa_q \sigma_q^2 \simeq 1\,.
  \label{eq:baseline}
\end{equation}
Since only the net charge is conserved, calculated as a difference
between the particle and the anti-particle contributions, the above
estimate is often referred to as the baseline by the
{\it Skellam distribution} that is the probability distribution of two
statistically independent variables.  It is also possible to apply the
Hadron Resonance Gas (HRG) model to estimate the baselines, and then,
$S_q\sigma_q$ and
$\kappa_q\sigma_q^2$ are generally suppressed by quantum statistical
effects, that reflects deviations of the Bose and the Fermi distribution
functions from the Boltzmann distribution.  The major strategy for the
QCD Critical Point search is to measure $S$ and $\kappa$ at various
$\sqrt{s_{NN}}$ and look for enhancement as compared to the
baseline~\eqref{eq:baseline}.
\vspace{1em}

\paragraph*{Baryon-rich Matter, an Approximate Triple Point, and Strangeness:}
From the HRG model estimate, the baryon number density along the
Chemical Freeze-out line is maximized around $T\simeq 150\MeV$ and
$\muB\simeq 400\MeV$
between $\sqrt{s_{NN}}=3-19.6\GeV$.  This has been experimentally
confirmed through the $K^+/\pi^+$ ratio peaked
around $\sqrt{s_{NN}}\simeq 8\GeV$.  It is intuitively easy to
understand that $K^+/\pi^+$ is such sensitive to the baryon density,
while $K^-/\pi^-$ is not.  In the heavy-ion collisions the time scale
is much shorter than the weak interaction, so the net strangeness
should be vanishing.  This means that the net chemical potential
coupled to $s$ quarks must be zero.  Since $s$ quarks have $S=-1$ and
$B=1/3$, the strangeness free condition leads to
\begin{equation}
  \mu_S \;\simeq\; \frac{1}{3} \muB \,,
\end{equation}
which means that the dense baryonic matter should contain strange
baryons or \textit{hyperons} which must be cancelled by mesons
involving $\bar{s}$ quarks such as $K^+$.  Therefore, $K^+$ is
enhanced at high density, while $K^-$ is not.

In this sense, this point reached around $\sqrt{s_{NN}}\simeq 8\GeV$ plays a
special role to tell us about a regime transition;  at smaller
$\muB\lesssim 400\MeV$ physics is dominated by mesonic degrees of
freedom, and at larger $\muB\gtrsim 400\MeV$ more baryons dominate.
Roughly speaking, the QGP transition is understood from the Hagedorn
transition with the exponentially rising meson-mass spectrum, while the the
transition at dense region arises from the Hagedorn transition with
the baryon-mass spectrum, and two Hagedorn transition lines cross just
around $\sqrt{s_{NN}}\simeq 8\GeV$.  In this way, the most baryonic
point around $\sqrt{s_{NN}}\simeq 8\GeV$ could be regarded as a
{\it QCD Triple Point} approximately facing baryon-less deconfined
matter, baryon-rich deconfined matter, and confined hadronic
matter~\cite{Andronic:2009gj}.

From the correlations between baryon number and strangeness, the QCD
Triple Point can be a landmark for the realization of the most strange
matter which contains hyperons.  In particular the interactions
between nucleons ($N$) and hyperons ($Y$), i.e., $Y$-$N$ and $Y$-$Y$
interactions are important parameters for the neutron star
structures.  The most strange matter would provide us with chances to
constrain those interactions.

\section*{CURRENT STATUS -- EXPERIMENT}

\begin{figure*}
  \includegraphics[width=0.75\textwidth]{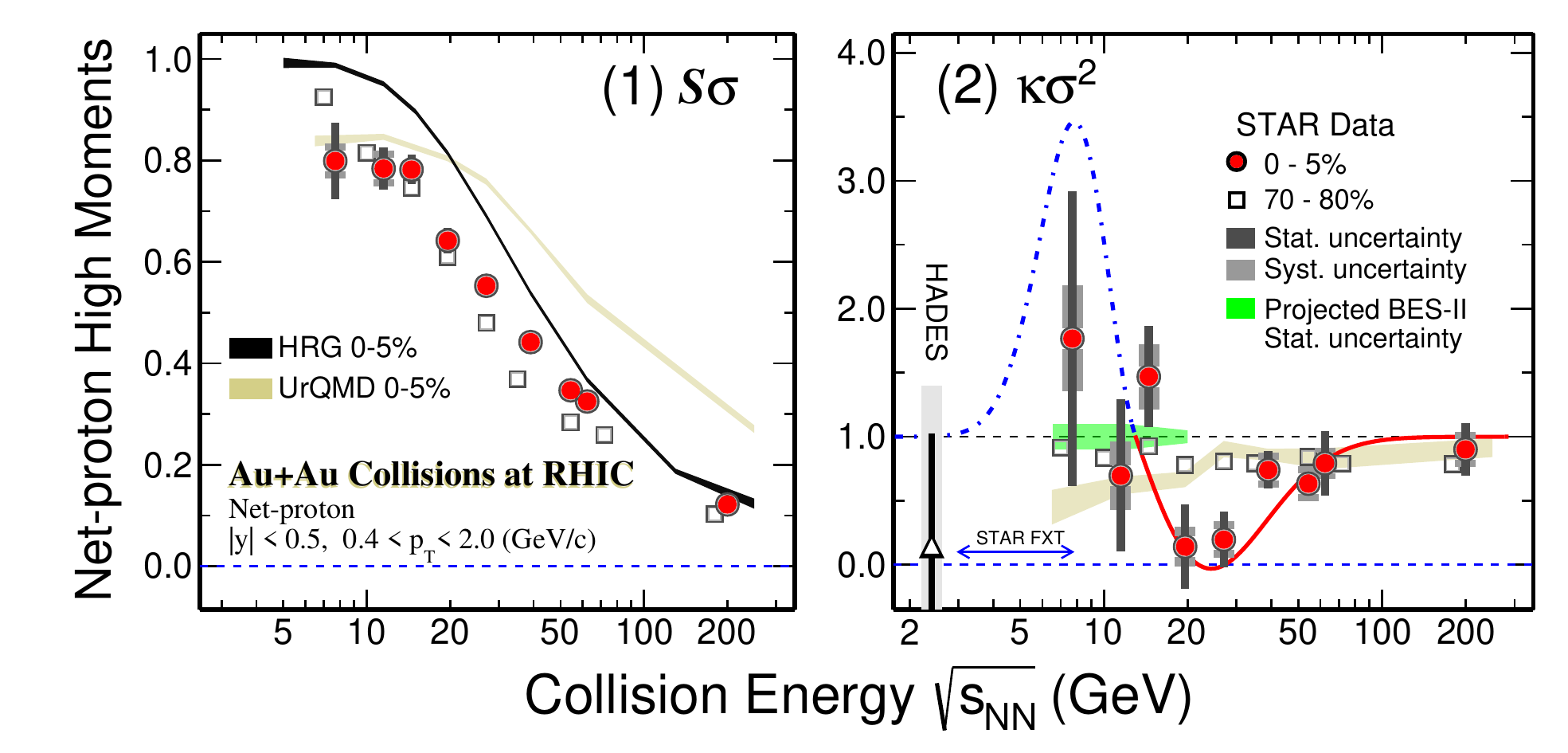}
  \caption 
  {Experimental results~\cite{Adam:2020unf} of $\sqrt{s_{NN}}$ dependence of 
    the net-proton
    $S\sigma$ (left) and $\kappa\sigma^2$ (right) from 70-80\% peripheral 
    (open squares) and 0-5\% central  (filled-circles) Au+Au 
    collisions. Projected statistical uncertainty for the second phase 
    of the RHIC BES program is shown by the 
    green-band. STAR experiments fixed-target program extends the center 
    of the mass collision energy down to 3 GeV. Results of calculations 
    are shown as black and gold bands for the HRG model and transport 
    model (UrQMD), respectively. The solid red  and the dashed
    blue line in (2) is a schematic representation of expectation from a
    QCD based model calculation in presence of a critical point.}
  \label{fig:Skappa}
\end{figure*}

\paragraph*{QCD Critical Point Search:}
The latest experimental results on the QCD Critical Point search are
shown in Fig.~\ref{fig:Skappa}.  The left panel (1) and the right
panel (2) show the measurements of $S\sigma$ and $\kappa\sigma^2$ for
the net-proton number distribution in Au+Au collisions at various
$\sqrt{s_{NN}}$.  We note that $N$ in Eq.~\eqref{eq:Nfluct} is the
baryon number, but neutrons are not electrically charged, and the
proton number is used experimentally as a proxy for the baryon
number.  Results are shown for both
central ($0-5\%$, small impact parameter) and peripheral ($70-80\%$,
large impact parameter) collisions.  Also shown are the expectations
from the HRG model and a transport based model called UrQMD, namely,
theories for central Au+Au collisions without including critical
fluctuations.

The following conclusions can be drawn: (a) As we go from lower order
moments ($S\sigma$) to higher order moments ($\kappa\sigma^2$)
deviations between central and peripheral collisions for the measured
values increases.  (b) Central $\kappa\sigma^2$ data show a
non-monotonic variation with collision energy at a significance of
$\sim$ 3$\sigma$~\cite{Adam:2020unf}.  (c) Experimental data
show deviation from heavy-ion collision models without a critical
point.  Although a non-monotonic variation of the experimental data
with collision energy looks promising from the point of view of the
QCD Critical Point search, a more robust conclusion can be derived
when the uncertainties get reduced and significance above $5\sigma$ is
reached.  The goal of the second phase of the BES program (BES-II) at
RHIC and the fixed traget (FXT) programs is to have high precision
measurements below $\sqrt{s_{NN}}= 3-19.6\GeV$.

The data presented in Fig.~\ref{fig:Skappa} provides the most relevant
measurements over the widest range in $\muB$ ($20-450 \MeV$)
to date for the critical point search, and for comparison with the
baryon number susceptibilities computed from QCD  to understand the
various features of the QCD phase structure.  The deviations of
$\kappa\sigma^2$ below the baseline~\eqref{eq:baseline} are
qualitatively  consistent with theoretical considerations including a
critical point~\cite{Stephanov:2011pb}.  However,
the conclusions on the experimental confirmation of the QCD Critical
Point might be made only after improving the precision of the
measurements at lower collision energies
and by comparing to the QCD calculations with critical point
behavior which includes the dynamics associated with heavy-ion
collisions.  See Ref.~\cite{Bluhm:2020mpc} for the latest report.
\vspace{1em}

\begin{figure}
  \includegraphics[width=0.9\columnwidth]{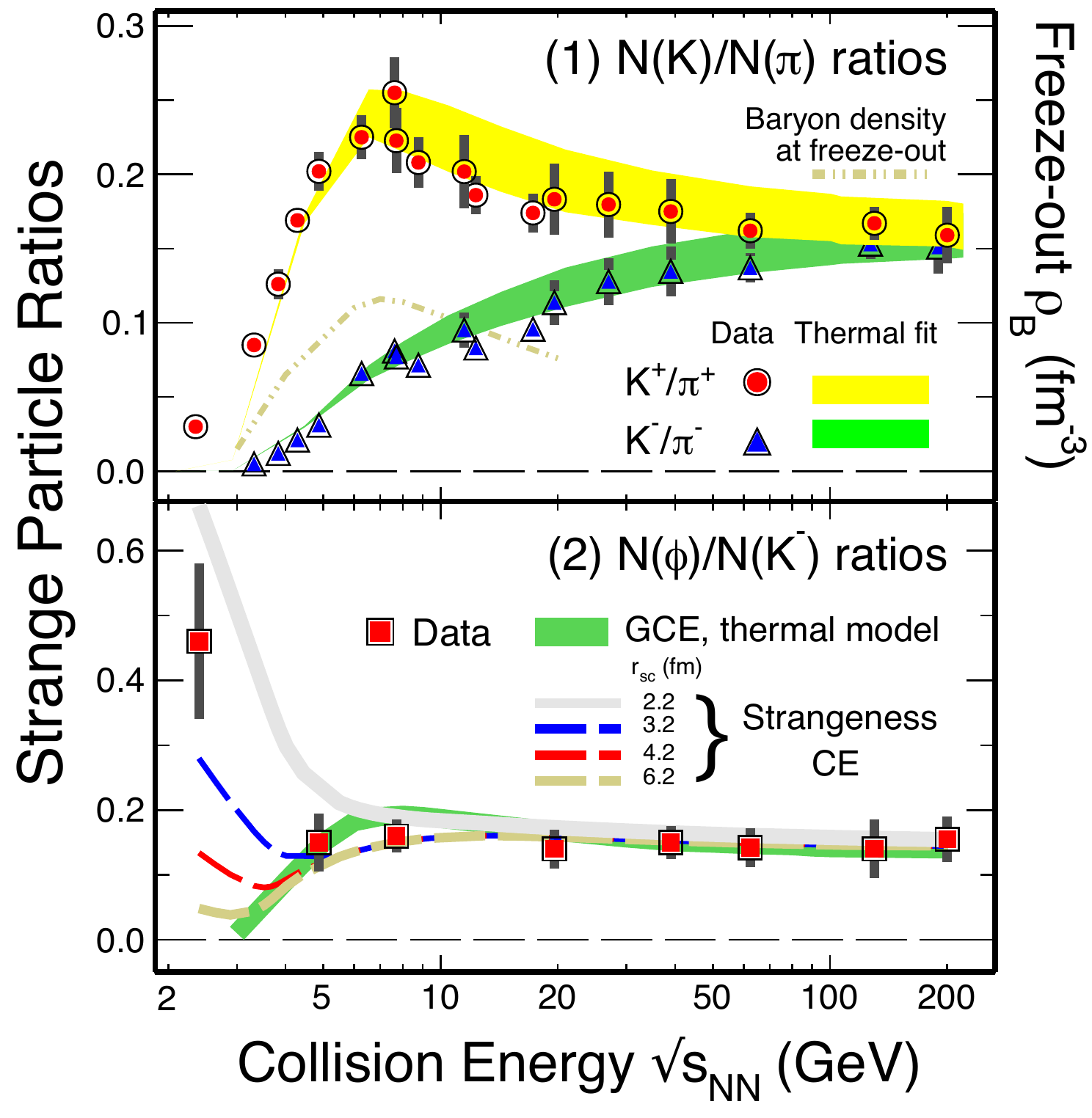}
  \caption 
  {(1) Particle yield ratios of kaon to pion as a function of $\sqrt{s_{NN}}$. Thermal 
    fits are also shown as bands in the plot. Dot-dashed line represents 
    the net-baryon density at the Chemical Freeze-out.  $K^+/\pi^+$
    (circles) trace the baryon density at the Chemical Freeze-out well, 
    while $K^-/\pi^-$ (triangles) increase smoothly as a function of 
    $\sqrt{s_{NN}}$.  (2) Particle yield ratios of $\phi$-meson to kaon 
    ($\phi/K^-$) as a function of $\sqrt{s_{NN}}$.  At energy below 
    8\,GeV the GCE fit no longer works, and the strangeness CE takes the 
    fit over.}
  \label{fig:Ktopi}
\end{figure}

\paragraph*{High Baryon Density Matter:}
Figure~\ref{fig:Ktopi}~(1) in the upper panel shows the energy
dependence of $K$/$\pi$ particle yield ratio.  The results are from
AGS~\cite{Ahle:1999uy,Ahle:1999va,Ahle:2000wq},
SPS~\cite{Afanasiev:2002mx,Alt:2007aa},
and RHIC~\cite{damczyk:2017iwn}.  These ratios reflect the strangeness
content relative to entropy of  the system formed in heavy-ion
collisions. The thermal model calculation is shown as yellow band for
$K^{+}$/$\pi^{+}$ and green band for $K^{-}$/$\pi^{-}$.  The
dot-dashed line represents the net-baryon density at the Chemical
Freeze-out as a function of collision energy, calculated from the
thermal model~\cite{Randrup:2006nr}.

The following observations can be made. (a) The collision energy 
dependence of both the ratios is fairly well described by a thermal
model calculation. (b) A peak position in energy dependence of 
$K^{+}$/$\pi^{+}$ is observed and has been suggested to be a signature
of a change in  degrees of freedom (baryon to
meson~\cite{Cleymans:2004hj} or hadrons to
QGP~\cite{Gazdzicki:1998vd}) while going from lower to higher
energies. (c) The calculated net baryon density exhibits a maximum as
the collision energy is scanned, with a value of about three-fourth of
the normal nuclear saturation density (i.e.,
$\rho_0\simeq 0.16\,\text{nucleons$/$fm$^3$}$). 
The collision energy where the maximum net-baryon density occurs is
very close to the peak position of the $K^{+}$/$\pi^{+}$ ratio. This
way of representation of the results from experimental measurement and
theory calculation serves to clearly demonstrate that the freeze-out
density and $K^{+}$/$\pi^{+}$ ratio could be related. (d) The
$K^{-}$/$\pi^{-}$ ratio seems unaffected by the changes in the
net-baryon density with collision energy and shows a smooth increasing
trend.

Through these measurements we have the knowledge of regions in collision
energy where the maximal net-baryon density is reached. This is an
important aspect in the context of planning of experiments that seek to
explore compressed baryonic matter.
\vspace{1em}

\paragraph*{Tests of Thermal Model -- GCE vs.\ CE:}

Relativistic statistical thermodynamics has been applied to systems
ranging from cosmology to heavy-ion collisions in laboratory. The
cosmological applications usually deal with systems having large volumes
and matter or radiation, hence the Grand Canonical Ensemble (GCE) is a
suitable description.
For heavy-ion collisions recorded in laboratory, the situation is
complicated due to the femtometer-scale nature of the systems. Often one
assumes (approximate) local thermal equilibrium for such
processes.  Further, such thermal models based on the GCE
employ chemical potentials to account for conservation of quantum
numbers on average. These GCE models have been able
to explain the particle production successfully for a wide range of
collision energies~\cite{Andronic:2017pug}.
However, conservation laws do impose restriction
on particle production if the available phase space is reduced.
Hence, the relativistic statistical thermodynamics 
provides two choices of the formalisms: a GCE and a canonical ensemble
(CE) approachs~\cite{Hagedorn:1984uy}. In the thermodynamic (large
volume) limit, the GCE and the CE formalisms are equivalent, but it is
an interesting question to ask where and when the transition from a
GCE picture to a CE one occurs for finite volume systems produced in
collisions at laboratory, where the collision
energy spans from a few GeV to a few TeV (three orders in magnitude).

Figure~\ref{fig:Ktopi}~(2) in the lower panel shows the energy
dependence of $\phi$/$K^{-}$ yield ratio. For most collision energies
the ratio remains constant. Similar to $K^{-}$/$\pi^{-}$ ratio, the
$\phi$/$K^{-}$ ratios seem not affected by the net-baryon
density. Below the collision energy where the freeze-out net-baryon
density  peaks [shown by the dot-dashed line in
Fig.~\ref{fig:Ktopi}~(1)] the  $\phi$/$K^{-}$ ratio starts to
increase. Thermal model calculations, adopting the GCE, which has been
quite successful in accounting for the
observed yields of the hadrons in heavy-ion collisions, explains the
measurements up to collision energy of $5\GeV$.  Then the GCE model
values decrease, while the increase in $\phi$/$K^{-}$ at lower
energies is explained by using a thermal model within the CE framework
for strangeness.
We note that a control parameter, $r_{\rm sc}$, is introduced for
strangeness CE results in Fig.~\ref{fig:Ktopi}~(2).  The physical
meaning of $r_{\rm sc}$ is a typical spatial size of $s\bar{s}$
correlations.  For smaller $r_{\rm sc}$, pairs of $s$ and $\bar{s}$ stick
together and the strangeness free condition is satisfied locally,
which suppresses the yield of $K^-$ and thus enhances $\phi/K^-$.
This makes a quantitative difference from the GCE results.
For a given volume of the whole system, $r_{\rm sc}$ determines how
close to the GCE/CE situation the strangeness sector in the system
should be.  Since $r_{\rm sc}$ reflects the intrinsic properties of
matter, the shifting from the GCE to the CE in strangeness signals a
considerable change of the medium properties.
Future measurements of $\phi$/$K^{-}$ at lower collision energies can
be used as an observable to estimate the volume in which the open
strangeness is produced (reflected by the value of $r_{\rm sc}$).
\vspace{1em}

\begin{figure}
  \includegraphics[width=\columnwidth]{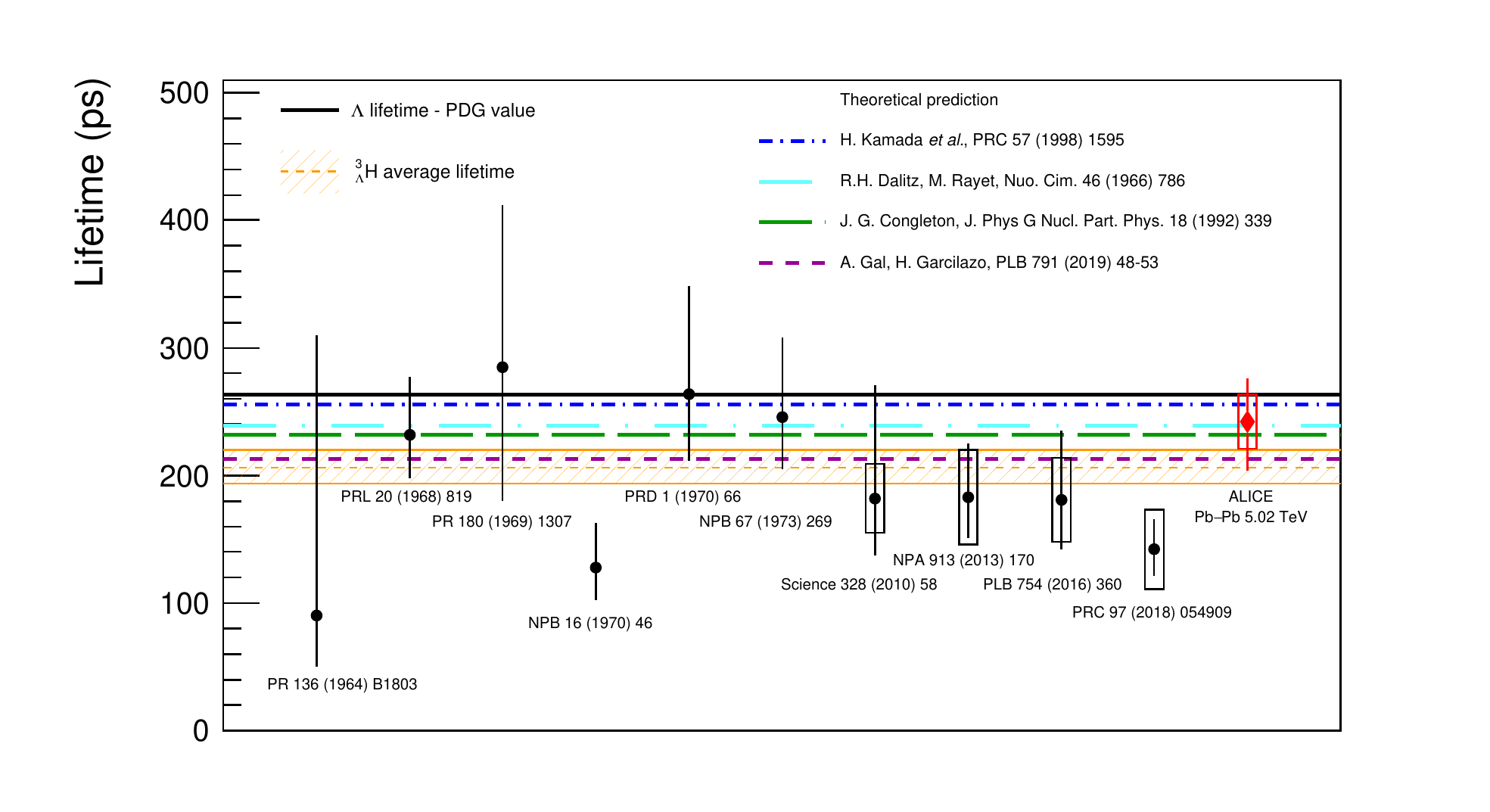}
  \caption{Collection of the hypertriton lifetime measurements from
    different experiments~\cite{Acharya:2019qcp,Adamczyk:2017buv}.
    The vertical lines and boxes are the statistical and the systematic
    uncertainties, respectively.  The orange band represents the
    average of the lifetime values and the lines at the edge
    correspond to 1$\sigma$ uncertainty.  The dot-dashed lines are four
    theoretical predictions.}
  \label{fig:hn-lifetime}
\end{figure}

\paragraph*{Lifetime of Hypernuclei:}
Hypernuclei are bound states of nucleons and hyperons, hence they are
natural hyperon-nucleon correlated systems~\cite{Abelev:2010rv}.
They can be used as an experimental probe to study the hyperon-nucleon
($Y$-$N$) interaction. Studying hypernuclei
properties is one of the best ways to investigate the strengths of
$Y$-$N$ interactions. Theoretically, the lifetime of a hypernucleus depends
on the strength of the $Y$-$N$ interactions. Therefore, a precise
determination of the lifetime of hypernuclei provides direct
information on the $Y$-$N$ interaction strength. The high energy
heavy-ion collisions at RHIC and LHC  create favorable conditions to produce
hypernuclei in significant quantities. At the moment, the experiments
have measured the production of the lightest hypernuclei, i.e., the
hypertriton, $\hypertriton$, which is a bound state of a proton, a
neutron and a $\Lambda$.

Figure~\ref{fig:hn-lifetime} shows a compilation of the measurement of the
hypertriton lifetime from various experiments and theory
calculations~\cite{Acharya:2019qcp,Adamczyk:2017buv}. The lifetime of
the (anti-) hypertriton is determined by reconstructing the mesonic
decay channels.  A statistical combination of all the experimental
results yields a global average lifetime of  $206_{-13}^{+15}$
picoseconds. The lifetime is about 22\% shorter than
the lifetime of a free $\Lambda$ of 263.2 $\pm$ 2.0 picoseconds,
indicating a possibility of a reasonable hyperon-nucleon interaction in
the hypernucleus system.
Most calculations predict the hypertriton lifetime to be in the
range of $213 - 256$ picoseconds. The $Y$-$N$
interaction  is  of  fundamental  interest, for it  controls the onset
of strange degrees of freedom in  high density nuclear  matter,
such as matter in the neutron star.
The lifetime measurements of hypernuclei thus provides a
crucial input for models attempting to understand physics of the
neutron star.
One should be aware of discrepancies in the measured lifetime of
$\hypertriton$ from RHIC and LHC{}. High statistics data are called
for in order to resolve these discrepancies.
\vspace{1em}

\paragraph*{Polarization and Spin Alignment:}
Recently, it was realized that the initial condition of the QGP
in relativistic heavy-ion collisions is subjected to two
extraordinary parameters:  the angular momentum and the magnetic field.
The angular momentum of the order of $10^{7}\hbar$ is theorized
to be imparted to the system through the torque generated when two
nuclei collide at non-zero impact parameter with center-of-mass
energies per nucleon of a few $100\GeV$~\cite{Deng:2016gyh}.
This leads to a thermal vorticity
of the order of $10^{21}$ per second for QCD matter formed in the
collisions~\cite{STAR:2017ckg}.  Further, when the two nuclei collide
in the LHC, an extremely strong magnetic field of the order of
$10^{15}$\,T is generated by the spectator protons, which pass by the
collision zone without breaking apart in inelastic collisions.  The
effect of the angular momentum (which is a conserved quantity) is
expected to be felt throughout the evolution of the system.
In contrast to that, the magnetic field
is transient in nature and stays for a short time of the order of
$\sim 0.1\fm/c$ unless the electric conductivity is large (but this is
disfavored, see discussions in Ref.~\cite{McLerran:2013hla}).
Just to give an idea of the magnitude of these values, the
highest angular momentum measured for nuclei (near the Yrast line) is
$\sim 70\hbar$ and the strongest magnetic field we have managed to
produce in the laboratory is $\sim 10^3$\,T using the electromagnetic
flux-compression technique. Getting experimental signatures of these
phenomena is not easy due to the femtoscopic nature of the system (both
in space and time) formed in the heavy-ion collisions.
Nevertheless, the experiments at RHIC and LHC have been able to
address this challenging problem.

It is known that the spin-orbit $LS$ coupling causes the fine
structure in atomic physics and the shell structure in nuclear
physics, and, is a key ingredient in the field of spintronics in
materials sciences. It is
also expected to affect the development of the rotating QGP created in
collisions of nuclei at high energies.
The extremely large initial value of the orbital angular momentum is
expected to lead to the polarization of quark spin
along the direction of the angular momentum of the
plasma's rotation due to the $LS$ coupling~\cite{Liang:2004ph}.
This should in turn cause the spins of vector (spin
= 1) mesons ($K^{*0}$ and $\phi$) to align~\cite{Acharya:2019vpe} and
hyperons like $\Lambda$ baryons to be 
polarized~\cite{STAR:2017ckg}. Both the hyperon polarization and the
spin alignment can be studied by measuring the angular distribution of
the decay products of $\Lambda$ and vector mesons. The hyperon
polarization is found to increase with decrease in heavy-ion collision
energy. The thermal vorticity values thus show that the QGP
formed in the collisions, along
with exhibiting the emergent properties of relativistic fluid, is also the
most vortical fluid found in nature~\cite{STAR:2017ckg}.
Meanwhile, the observed spin alignment of vector mesons (with $J=1$)
was quantified by obtaining the probability of finding a vector meson
in a $J_z=0$ state along the $z$ direction that is the direction of
the orbital angular momentum of the rotating QGP{}.
The momentum dependence of these probability values indicated
polarization of quarks in the presence of large initial angular
momentum in heavy-ion collisions and a subsequent hadronization by the
process of recombination~\cite{Acharya:2019vpe}.

\section*{FUTURE DIRECTIONS -- THEORY}
There remain a lot of theoretical challenges in understanding physics
of dense baryonic/quark matter with magnetic field and rotation.  We
give brief discussions on some topics in order.
\vspace{1em}

\paragraph*{QCD Phase Structures and Quark Matter at High Baryon Density:}
Theoretically, it is highly nontrivial how quarks can melt
from hadrons in cold and dense matter.  Unlike hot QCD even an
approximate measure for quark deconfinement is still unknown or such
an order parameter simply may not exist.

The QCD Critical Point is a landmark on the QCD phase diagram and the
next intriguing question is where we can find quark matter.  One might
naively think that the asymptotic freedom with a large quark chemical
potential, $\muq\gg\LQCD$, makes quarks unbound from hadrons, but this
is not necessarily true.  When $\muq$ is large, quarks form a Fermi sphere,
and the typical energy scale of quarks near the Fermi surface is
$\sim\muq$.  However, gluons can still carry soft momenta, mediating
confining forces.  Therefore, excitations on top of the Fermi
surface are still confined, while the Fermi sphere itself is dominated by
quarks, and this refined picture of a dense baryonic state is called Quarkyonic
Matter~\cite{McLerran:2007qj}.

One can develop a more precise definition of Quarkyonic Matter by
deforming the fundamental theory;  in reality $\Nc=3$ where
$\Nc$ is the number of colors, and one can significantly simplify
theoretical treatments by taking the $\Nc\to\infty$ limit.  In this
special limit the
ground state could have an inhomogeneous crystalline
shape~\cite{Kojo:2009ha} (see also a review~\cite{Buballa:2014tba} for
comprehensive studies of inhomogeneous phases).  In reality mesonic
fluctuations would destroy inhomogeneity, but some remnant
correlations can still remain.
Those remnants of enhanced spatial correlations would
increase the density fluctuation.  For experimental detections to
discriminate it from bubble formation associated with a first-order
transition beyond the QCD Critical Point, more theoretical works are
needed.
\vspace{1em}

\paragraph*{Neutron Star Phenomenology:}
We specifically pick two problems here in neutron star phenomenology.
One is a question of whether quark matter is found in cores of the
neutron star, which is a continued subject from the above problem of
the phase diagram, and the other is what is called the hyperon puzzle.

It is the experimental fact that massive neutron stars whose masses
are greater than $2M_{\odot}$ exist, where $M_{\odot}$ represents the solar
mass.  This observations is strong
enough to constrain the stiffness of the equation of state (EoS) and a
strong first-order phase transition has already been excluded.   Thus,
even if quark matter existed in cores of the neutron star, it is
likely that there is only a smooth crossover or a weak first-order
transition from nuclear to quark matter.

Matter created in the heavy-ion collision is regarded better as hot
and dense baryonic matter.  Since the physical observables are
sensitive to the EoS, the global analysis of experimental data would
quantify the most likely regions of EoS parameters.  Such a program of
the global Bayesian analysis has already been successful at large
$\sqrt{s_{NN}}$ where the experimentally inferred EoS is found to be
consistent with the lattice-QCD results~\cite{Pratt:2015zsa}.  The
same machinery would in principle constrain the EoS of hot and dense
baryonic matter.  One of the most interesting EoS parameters is the
speed of sound $c_s^2$, which hints the presence of quark matter as
discussed in Ref.~\cite{Annala:2019puf}.
We also mention that the global analysis of the flow measurements
could constrain the viscosities of dense matter (apart from leptonic
contributions), which should be useful for considerations of the
$r$-mode evolutions of the neutron star~\cite{Andersson:2000mf}.  See,
for example, Ref.~\cite{Kolomeitsev:2014gfa} for a theoretical
estimate of viscosities of dense nuclear matter.

Let us turn into the hyperon puzzle that has twofold manifestations.
If the baryon density reaches several times $\rho_0$, inside the
neutron star to balance the gravitational force, it is
energetically more favorable to activate the strangeness degrees of
freedom.  One problem is that the introduction of strangeness
generally softens the EoS and it would become more
difficult to support the neutron stars
with the mass $\gtrsim 2M_\odot$.  Another problem is, once hyperons
are favored, the direct Urca process would shorten the time scale of
the neutron star cooling, which makes the neutron star too cold.
Thus, the threshold of the neutron star mass to open the direct Urca
process is an important parameter, and this is dictated by the $Y$-$N$
and $Y$-$Y$ interactions as well as three-body forces involving hyperons.

Theoretically speaking, the most promising approach is the
first-principles calculation of the baryon interactions including
nontrivial strangeness from the lattice-QCD
simulation~\cite{Nemura:2017vjc}.  In the HAL QCD method the
Nambu-Bethe-Salpeter wave functions are computed on the lattice, from
which the potential is extracted, see Ref.~\cite{Aoki:2020bew} for a
review on the HAL QCD method including hyperon results.  For example,
$p\Xi^-$ correlation has been theoretically predicted to have attractive
interaction~\cite{Hatsuda:2017uxk}; this is an interesting system
since $\Xi^-\sim dss$ is a multi-strange baryon, and the experimental
signature is reported~\cite{Acharya:2019sms}.  Also, the correlations of
$\Omega\Omega$ and $N\Omega$ have also been estimated in
Ref.~\cite{Morita:2019rph} based on the lattice-QCD determined
potential.  For more comprehensive discussions to quantify the
potential from the correlations in heavy-ion collisions, see a recent
review~\cite{Cho:2017dcy}.
\vspace{1em}

\paragraph*{Dibaryons and Diquarks:}
$\Omega\Omega$ is an interesting candidate for one of possible
dibaryons~\cite{Gongyo:2017fjb} which are six-quark objects.  There is
a long history of the dibaryon hunting (see
Ref.~\cite{Clement:2016vnl} for a review);  the idea is traced back to the
conjecture on the $H$-dibaryon~\cite{Jaffe:1976yi}.  One might
think that the deuteron is also a six-quark bound state, but what is
special about the $H$-dibaryon is that the diquark correlation
plays an essential role.  From the one-gluon exchange interaction the
color-triplet diquarks are favored, and the low-energy reduction leads
to the Breit interaction involving the color and the spin degrees of
freedom.  It is an established notion that the energetically most
favored channel is the spin-singlet and the flavor-triplet, and
diquarks in this channel are called ``good diquarks'', while the
second stable diquarks, i.e., ``bad diquark'' are found in the
spin-triplet and the flavor-sextet channel.  The structure of the
$H$-dibaryon is considered to be dominated by three good
diquarks, i.e., $H\sim (ud)(ds)(su)$.

It is still challenging to find a direct signature of the
strong diquark correlation.  From the theoretical point of view, the
difficulty lies in the fact that diquarks are not gauge invariant.
Nevertheless, the density-density correlation in baryon wave-functions
could quantify the diquark correlation in a gauge-invariant
way~\cite{Alexandrou:2006cq}.
Interestingly, the diquark correlations would be more prominent at
higher baryon density.  Actually, it is a solid theoretical prediction
that QCD matter at asymptotically high density should be a color
superconductor in which the diquarks form condensates.  If there is no
sharp transition separating baryonic matter from
color-superconducting quark matter, as is conjectured in the
quark-hadron continuity scenario, one can expect some remnants of
the diquark correlations in density regions accessible by the
heavy-ion collision.  The interesting question is whether diquarks are
treated as active thermal degrees of freedom, participating in the
thermal model in dense matter, see Ref.~\cite{Shuryak:2004tx} for a
model with colored thermal excitations like diquarks.  Since the
lattice-QCD calculation is not functional at finite density, the test
can be made only in comparison to experimental data.
\vspace{1em}

\paragraph*{Femto-Nova Rotating with Magnetic Fields:}
The profound feature of matter created in the heavy-ion collision is
that non-central collisions are accompanied by vorticity and magnetic
fields as illustrated in Fig.~\ref{fig:FemtoNova}.  Such a hot, dense,
and rotating object exposed under the magnetic field can be found as
an emulator of the proto-neutron star after the supernova, and we may
well call this heavy-ion system the Femto-Nova.

The Femto-Nova investigations have a lot of research potentials.
Relativistic rotation and magnetic fields would change the properties
of matter, or even the phase diagram should be
affected~\cite{Jiang:2016wvv}.  In numerical simulations of the
supernovae and the neutron-star-merger, such effects of rotation and
magnetic fields are not taken into account yet.  Thus, the heavy-ion
collision experiments can constrain uncertainties in the interplay of
rotation and magnetic field in strongly interacting matter.  In
particular, relativistic formulations of spin- and
magneto-hydrodynamics are still in the middle way of developments.

Rotation and magnetic fields are of paramount importance also from the
point of view of the topological effects.  Once the density (finite
$\muB$), the rotation (finite angular velocity $\bomega$), and the
magnetic field $\bB$ are coupled together, the theory tells us that the
\textit{chiral seperation effect} (CSE) and the
\textit{chiral vortical effect} (CVE)
should appear (see Ref.~\cite{Kharzeev:2015znc} for a review):
\begin{equation}
  \boldsymbol{J}_5 = \sigma_s \bB\,,\qquad
  \boldsymbol{J}_5 = \sigma_v \bomega\,,
  \label{eq:topoJ}
\end{equation}
where $\sigma_s \propto \muB$ if the particle masses are negligible.
The coefficient $\sigma_v$ has two components;  one is $\propto T^2$ and
the other $\propto \muB^2$.  Here, $\boldsymbol{J}_5$ is the axial
current, and its physical meaning is the spin expectation value of
matter.  Therefore, the first in Eq.~\eqref{eq:topoJ} physically
represents the spin polarization, and what is nontrivial in
relativistic systems is that the spin and the momentum of massless
fermions are tightly correlated with a certain handedness.
Therefore, the global spin polarization results from the CSE leading
to the chirality separation associated with a chirality flow along the
polarization.

The second equation, i.e., the CVE, looks like a counterpart of the
CSE with $\bB$ replaced by $\bomega$, but physical interpretations are
rather nontrivial.  In this context the physical meaning of the CVE is the
relativistic realization of the \textit{Barnett effect} (see
Ref.~\cite{Barnett1935} for a review by Barnett himself);  a
mechanical rotation yields nonzero magnetization~\cite{Gao:2012ix}.
One might then wonder if the chiral anomaly mechanism could be an
independent origin from the conventional $LS$ coupling.  Actually, in
the nonrelativistic Barnett effect, the magnetization is inversely
proportional to the gyromagnetic ratio, and so it is proportional to
the mass;  this has motivated the nuclear Barnett effect
experiment~\cite{PhysRevLett.122.177202}.  The CVE is more
prominent, however, for massless fermions, and their mass dependences
look competing.  The theoretical framework is not yet
complete to incorporate all those effects consistently for arbitrary masses of
fermions.  Establishing a firm bridge between nonrelativistic and
relativistic (as seen in the heavy-ion collision) Barnett effects is a
challenging subject in theory.
\vspace{1em}

\section*{FUTURE DIRECTIONS -- EXPERIMENT}

\paragraph*{More on the QCD Critical Point:}
As discussed above, the search for the QCD
Critical Point has been led by the RHIC BES program,
where the collision energy has been dialed down from $200\GeV$
(see Fig.~\ref{fig:Skappa}). It spans a $\muB$-range from $20$ to
$400\MeV$ of the phase diagram.
The fluctuations near the QCD Critical Point are
predicted to make $\kappa\sigma^2$ swing below its baseline value
($=1.0$) as the critical point is approached, then going well above,
with both the dip and the rise being significant for
head-on nuclear collisions~\cite{Stephanov:2011pb}.
The data show a substantial drop and intriguing hints of
a rise for the lowest energy collisions, although the uncertainties at
present are too large to draw definitive conclusions [see panel (2) of
Fig. ~\ref{fig:Skappa}]. The ongoing phase-II of the BES
program and the fixed target program at RHIC aim to
gather high statistics data to look for this important landmark
in the QCD phase diagram.  The lattice-QCD calculations suggest
that the QCD Critical Point, if exists, lies for
$\muB/T\gtrsim 2$~\cite{Bazavov:2017dus}. Thus, the
role of upcoming high baryon density experiments as listed below
becomes important in the critical point search program. Not only they
extend the search to high $\muB$ regions ($\approx 750\MeV$) of
the phase diagram, they also
provide a reverse approach of studying the critical point observable
by dialing up the beam energy. This approach is complimentary to the
current searches and the observable studied from both directions of
collision energy (i.e., from both above and below the QCD Critical
Point)  is expected to meet at a common point.
This will complete test of the theoretical prediction of non-monotonic
variation of $\kappa\sigma^2$ with $\sqrt{s_{NN}}$.
\vspace{1em}

\begin{figure}
\includegraphics[width=0.9\columnwidth]{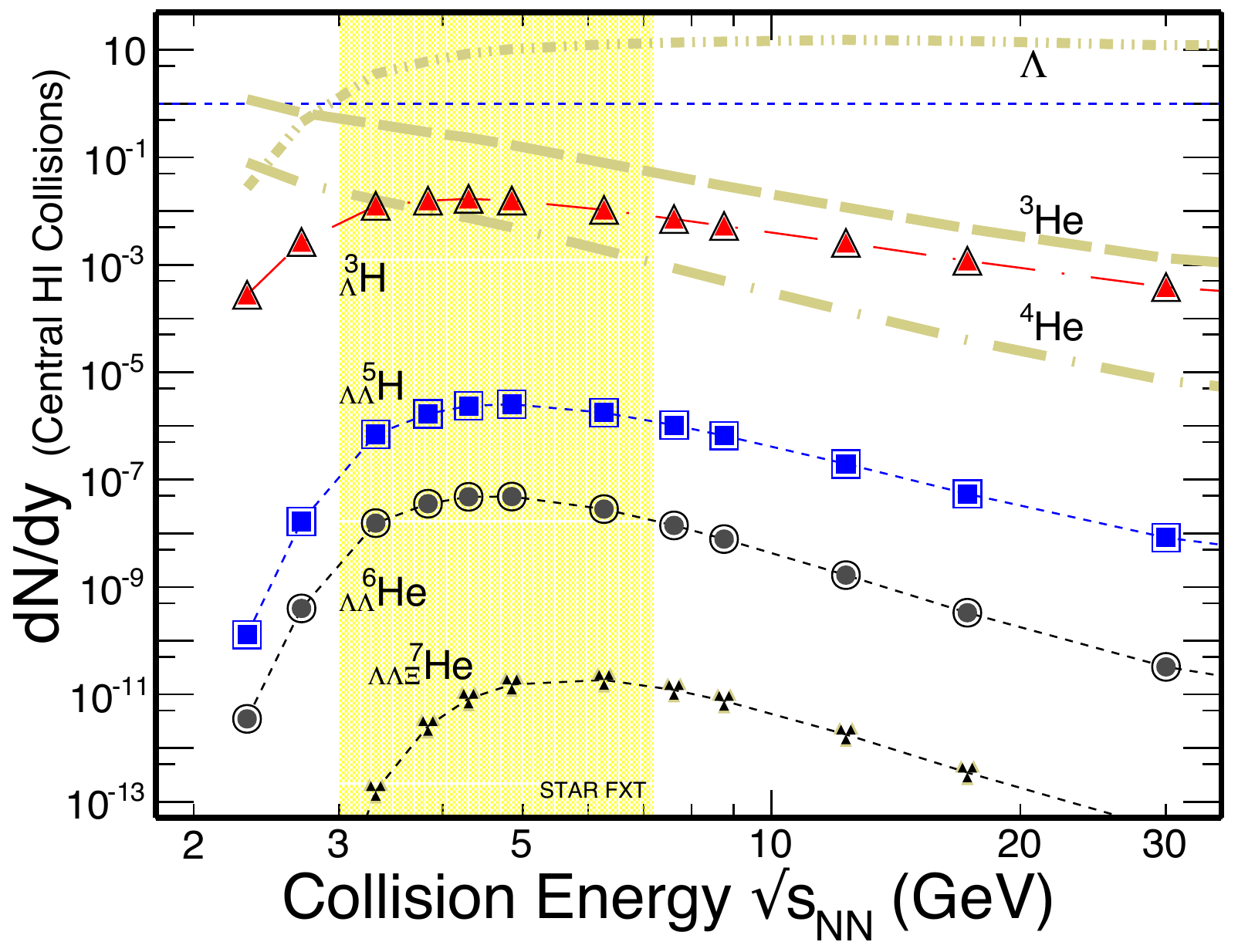}
\caption{Heavy nuclei (dashed lines) and (multi-)hypernuclei (symbols)
  yields calculated in the HRG
  model~\cite{Andronic:2005yp,Andronic:2010qu} for central heavy-ion
  collisions at mid-rapidity as a function of center-of-mass
  energy. Also shown for comparison are the yields of $\Lambda$
  hyperon from the same model. The collision energy region of fixed
  target program in STAR is indicated as a yellow band.  Model results
  from Ref.~\cite{Andronic:2010qu}.}
\label{fig:hyper}
\end{figure}

\paragraph*{Light Hypernuclei Production:}
Figure~\ref{fig:hyper} shows the mid-rapidity yields of light-nuclei
and (multi-)hypernuclei, from thermal (HRG) model calculations, shown as a
function of colliding energy.  All of the data points are from
Refs.~\cite{Andronic:2005yp,Andronic:2010qu} (see also
Ref.~\cite{Steinheimer:2012tb} for a theoretical analysis).  As one
can see in the figure, all of the light hypernuclei yields are peaked
around $3 - 8\GeV$ range fully covered by both STAR fixed target (FXT)
program~\cite{bes2} (hatched region) and future CBM experiment at
FAIR~\cite{Ablyazimov:2017guv}.

Data of $K^+$ over pion ratios show a peak at the center-of-mass
energy of $8\GeV$ implying that the baryon density at Chemical
Freeze-out reaches maximum around this colliding energy,
see Fig.~\ref{fig:Ktopi}~(1) at top panel.  Due
to the relatively low production threshold, the production of the
$\Lambda$ hyperon becomes abundant.  The coalescence
process~\cite{Baltz:1993jh} combines these advantages and leads to the
copious production of the hypernuclei in this energy region.  The
strangeness degrees of freedom is therefore introduced into
the dense nuclear matter.
The cross sections for hypernuclei productions
in high energy nuclear collisions are much higher than that in either
elementary collisions or Kaon induced interactions, making the
heavy-ion collision as a hypernuclei factory (HNF).   The HNF offers a
great opportunity for
studying fundamental interactions of hyperon-nucleon ($Y$-$N$),
hyperon-hyperon ($Y$-$Y$) within the many-body baryonic system and the
spectroscopy of nuclear structure with
strangeness~\cite{Hashimoto:2006aw}.  In addition, these nuclear
collisions provide the means to study the inner dynamics of compact
stars in the laboratory.  We should note that most of the studies on
hypernuclei so far utilized the ``light system'' with electron or
pion or Kaon beams.  In such cases the hypernuclei were produced in
vacuum.  Data on hyperon production in nuclear collisions is
scarce~\cite{Rappold:2013jta}.  Measurements of hypernuclei
collectivity in the truly heavy-ion, Au+Au, collisions, for example,
allow one to extract information on the transport properties (crucial
for the neutron star stability, see Ref.~\cite{Kolomeitsev:2014gfa})
as well as the $Y$-$N$ interaction driven
EoS, with the strangeness degrees of freedom, in the hot and dense
environment where the baryon density could be very high.  Simulations
for neutron star inner properties crucially
depend on the EoS, see Ref.~\cite{Vidana:2000ew} for the effect of
$Y$-$Y$ interactions, and also Ref.~\cite{Maruyama:2007ey} for the
hyperon effects including a possibility of quark mixture.  Furthermore, 
an additional benefit of the unique high baryon density environment is
the enhanced production of multi-$\Lambda$ hypernuclei as already
suggested in Fig.~\ref{fig:hyper}.

The future accelerator based experiments, as introduced below and
aimed for high baryon density matter, compresses the baryonic matter
in heavy-ion collisions around $2-8\GeV$, which takes place in an
ideal location for serving as the HNF as shown in Fig.~\ref{fig:hyper}.
Hence, these future experiments can make tremendous contributions
towards detecting and measuring the yields of hypernuclei and their 
life-time. This will then provide valuable inputs to understanding the 
hyperon-nucleon ($Y$-$N$) interactions in heavy-ion collisions and
inner dynamics of the compact stars.
\vspace{1em}

\begin{figure}
\includegraphics[width=0.95\columnwidth]{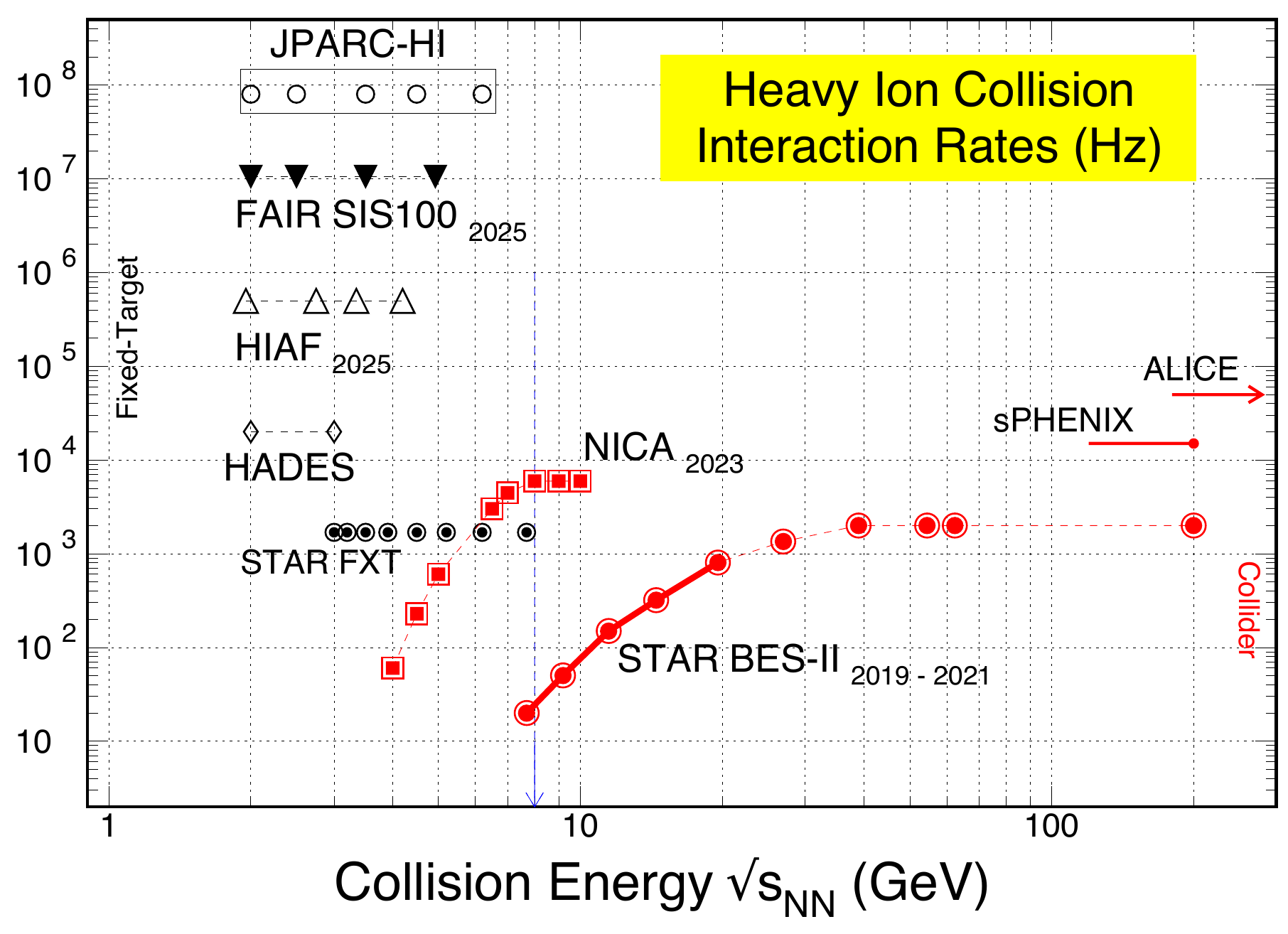}
\caption{Interaction rates (in Hz) for high-energy nuclear collision
  facilities. Collider mode: the second phase RHIC beam energy scan
  (BES-II)~\cite{bes2} for $7.7\GeV <\sqrt{s_{NN}} <19.6 \GeV$
  (filled-red-circles) and NICA
  (filled-red-squares)~\cite{Geraksiev:2019fon}. Fixed target mode:
  STAR fixed target (FXT) program for $3.0\GeV<\sqrt{s_{NN}}< 7.2 \GeV$
  (filled-black-circles), FAIR (CBM, SIS)~\cite{Ablyazimov:2017guv},
  HADES~\cite{Agakishiev:2009am},  J-PARC~\cite{jparc}, and
  HIAF~\cite{Ruan:2018fpo}.  Also shown for reference the rates of
  ALICE at LHC~\cite{Dainese:2019rgk} and sPHENIX at
  RHIC~\cite{Roland:2019cwl}.}
\label{fig:rates}
\end{figure}

\paragraph*{Fluid Vorticity of High Baryon Density Matter:}
Experiments at RHIC and LHC have observed that the polarization of
hyperons and vector mesons have a distinct energy dependence. Their
values increase with decrease in collision energy. The physics
reasons attributed for the observed energy dependence are twofolds.
Firstly, the baryon stopping is enhanced and shear flow patterns in the beam
direction emerge, and secondly, a shorter lifetime of the fluid phase thereby
allows perseverance of the initial vorticity in the system from
getting diluted~\cite{Karpenko:2016jyx}. The possibility of high
interaction rate experiments in high baryon density matter at the
upcoming facilities, opens up
an unique facility to study relativistic interplays of the spin, the
orbital angular momentum, and the magnetic field in QCD matter.
This will guide theoretical developments in the field of
relativistic spin- and magneto-hydrodynamics.
\vspace{1em}

\paragraph*{Future Experimental Facilities for High Baryon Density Matter:}
The upcoming facilities for studying high baryon density matter
includes
(a) Nuclotron-based Ion Collider fAcility (NICA) at the Joint Institute
for Nuclear Research (JINR), Dubna, Russia,(
(b) Compressed Baryonic Matter (CBM) at Facility for  
Antiproton and Ion Research (FAIR), Darmstadt, Germany,  
(c) Japan Proton Accelerator Research Complex (J-PARC), Ibaraki, Japan,
and (d) CSR External-target Experiment (CEE) at
High Intensity heavy-ion Accelerator Facility (HIAF)~\cite{hiaf1},
Huizhou, China.  The interaction
rates of these upcoming experiments compared to existing and other
future facilities are shown in Fig.~\ref{fig:rates}.
Note that all the new facilities under construction are focused in the
energy region where the baryon density is high.  Below we discuss
briefly the salient features of these four experiments.
\vspace{0.5em}

\underline{\it NICA\atmark JINR}: At this new accelerator complex
under construction the plan is to provide accelerated particle beams
both in collider (Multi Purpose Detector) and fixed target (Baryonic
Matter at Nuclotron) modes~\cite{Geraksiev:2019fon}.
The gold nuclei collision energies will be in the range,
$\sqrt{s_{NN}}= 4-11\GeV$.  The physics goals are dominantly to
explore the QCD phase diagram through measurements of particle yields,
collective flow, femtoscopy etc. In addition, it also emphasizes
studying polarization of hyperons and investigating hyperon-nucleon
($Y$-$N$) interactions through hypernuclei production.
As one sees in Fig.~\ref{fig:rates}, NICA connects the high-energy
collider experiments with the FXT experiments nicely.
\vspace{0.5em}

\underline{\it CBM\atmark FAIR}:  This facility currently under
construction will offer the opportunity to study nuclear collisions at
extreme interaction rates.  It will initially comprise of the SIS100
ring which provides energies for gold beams of
$\sqrt{s_{NN}} = 2.7 - 4.9 \GeV$ and $\muB >500\MeV$.  The CBM
detector at FAIR has been designed as a multi-purpose device which
will be capable to measure hadrons, electrons, and muons in heavy-ion
collisions over the above beam energy range at interaction rates up to
10\,MHz for selected observables. The physics goals include studying
the phase structure of the QCD phase diagram (i.e., the order of the
transition, the QCD Critical Point, and chiral symmetry), possible
modification of properties of hadrons in dense baryonic matter, and
the EoS at high density as is expected to be relevant to the cores of
neutron stars through measurements of hypernuclei and heavy
multi-strange objects.
\vspace{0.5em}

\underline{\it JPARC-HI\atmark KEK/JAEA}: The idea of this facility is
under discussions for several years and the planned J-PARC-HI will
provide heavy-ion beams up to uranium for center-of-mass energies of
$2-6.2\GeV$.  This corresponds to exploring the QCD
phase diagram in very high baryon densities~\cite{jparc}.
It excepts to carry out important measurements
including dileptons to understand QCD transitions,
in-medium modifications of $\rho$, $\omega$, and $\phi$ mesons
decaying into dileptons, rare particles such as
multi-strangeness hadrons, exotic hadrons, and hypernuclei utilizing
high rate beams.
According to the plan, this will be the experiment with the highest
beam rate capability up to $100$MHz allowing precision measurements
for rare processes in heavy-ion collisions.
\vspace{0.5em}

\underline{\it CEE\atmark HIAF}:
The complex of the HIAF is under construction and it is expected to be
in operation in 2025. The machine is designed to deliver bright ion
beams of protons and heavy nuclei such as uranium with the
center-of-mass energy up to $10\GeV$ and $4\GeV$, respectively. A
superconducting dipole
magnet spectrometer experiment (CEE)~\cite{Lu:2016htm} is also under
construction.  In many respects, this is a simple hadron spectrometer 
with the main physics focused on the measurements of (i) proton, light
nuclei including hypernuclear production and correlation for
understanding the QCD phase structure and (ii) meson ratios for
extracting the EoS at the high baryon density region.
\vspace{0.5em}

Future new experiments are all designed with high rates, large
acceptance, and the-state-of-the-art particle identification,
at the energy region where baryon density is high, i.e.,
$500\MeV < \muB < 800\MeV$, see Fig.~\ref{fig:rates}.

\section*{CONCLUSIONS}

We reviewed what we have understood so far and what we are trying to
understand in the future using the relativistic heavy-ion collisions.
There are three important physics targets:
\begin{itemize}
  \item[(1)] Scanning the QCD phase diagram and seeking for the QCD
    Critical Point.
  \item [(2)] Constraining the $Y$-$N$ and $Y$-$Y$ interactions and the EoS
    in dense baryonic matter including strangeness degrees of freedom.
  \item[(3)] Exploring the effects of large angular momentum and strong
    magnetic fields.
\end{itemize}
For (1) ``Criticality'' is essential to detect the critical point, and
the extraction of the EoS as (2) needs the global analysis including
``Collectivity'', and the physics of (3) exhibits topologically
nontrivial effects once nonzero ``Chirality'' is involved.  These three
C's abbreviate the future directions of the heavy-ion
collision physics.

The first-principles calculations from the lattice-QCD simulation have
shown tremendous progresses with the cutting-edge computing
technologies also toward the high-density region.  Now understanding
the QCD phase structure is in need of experimental guides together
with theoretical approaches.  The QCD Critical
Point is a landmark, and the next question is what awaits beyond it.
If the first-order phase transition is reached, the spinodal
decomposition and the nucleation processes would lead to
characteristic patterns of baryon fluctuations.

In the heavy-ion collisions of center-of-mass energy below $15\GeV$,
one of the important features is that the
baryon density is high enough to be above the threshold for the
strangeness production.  The interesting observation is transitional
behavior from the grand canonical to the canonical ensembles in the
strangeness sector with different collision energies.  Also, with
abundant strangeness, $N$-$N$, $Y$-$N$, and $Y$-$Y$
interactions can be investigated not only in the vacuum but in an
environment with surrounding baryonic mean field.  Under such a
situation, measurements of baryon correlations and collectivity
involving multi-strange hadrons such as
$\phi$-meson, $\Lambda$, $\Xi$, and $\Omega$-baryons, and hypernuclei
will provide us with the information on the
EoS relevant to the neutron star structures and simulations of the
supernovae and the neutron star merger.  Although it is not covered in
this article, as long as the penetrating observables are concerned,
dilepton mass distributions for example give us informaion on the
initial thermal properties for matter created in the heavy-ion
collision (see a recent review~\cite{Salabura:2020tou}).

The unique property of matter in the heavy-ion collision is the
presence of rotation (or the angular momentum) and the external
magnetic fields.  Such extreme environments at high baryon density,
rapid rotation, and strong magnetic fields can be found not only in
the heavy-ion collisions but also in astrophysical phenomena.
Therefore, revealing those effects in the controlled laboratory
experiments shall bring conner stones for understanding the nature of
visible matter, through the Femto-Nova, in the Universe.

\section*{Acknowledgments}
The authors thank Drs.\
A.~Andronic,
A.~Bamba,
N.~Herrmann,
K.~Redlich,
H.~Sako,
S.~Samanta
for exciting discussions.
The authors also thank the colleagues from STAR and ALICE 
collaborations. 
K.F.\ was supported in part by Japan Society for the Promotion of 
Science (JSPS) KAKENHI, Nos.\ 18H01211 and 19K21874. 
B.M.\ was supported in part by the Chinese Academy of Sciences  
President's International Fellowship Initiative
and J C Bose Fellowship from
Department of Science of Technology, Government of India.
N.X.\ was supported in part by
the National Science Foundation of China, No.\ 11927901.

\bibliographystyle{apsrev4-1}
\bibliography{review}
\end{document}